\documentclass[12pt]{article}%
\usepackage{amssymb}
\usepackage{amsmath}
\usepackage{amsfonts}
\usepackage{graphicx}
\usepackage{xcolor}%
\providecommand{\U}[1]{\protect\rule{.1in}{.1in}}

\begin{document}
\bigskip\begin{titlepage}
		\vspace{.3cm} \vspace{1cm}
		\begin{center}
			\baselineskip=16pt \centerline{\bf{\Large{ Mimetic Ho\v{r}ava Gravity}%
			}}
			\vspace{1truecm}
			\centerline{\large\bf Ali H.
				Chamseddine$^{1,2}$\ , Viatcheslav Mukhanov$^{2,3,4}$\ , Tobias B. Russ$^{2}%
				$\ \ } \vspace{.5truecm}
			\emph{\centerline{$^{1}%
					$Physics Department, American University of Beirut, Lebanon}}
			\emph{\centerline{$^{2}%
					$Theoretical Physics, Ludwig Maxmillians University,Theresienstr. 37, 80333 Munich, Germany }%
			}
			\emph{\centerline{$^{3}%
					$MPI for Physics, Foehringer Ring, 6, 80850, Munich, Germany}}
			\emph{\centerline{$^{4}%
					$School of Physics, Korea Institute for Advanced Study, Seoul 02455, Korea}}
		\end{center}
		\vspace{2cm}
		\begin{center}
			{\bf Abstract}
		\end{center}
	We show that the scalar field of mimetic gravity could be used to construct
	diffeomorphism invariant models that reduce to Ho\v{r}ava gravity in the
	synchronous gauge. The gradient of the mimetic field provides a timelike unit vector field that allows to define a 			projection operator of four-dimensional tensors to three-dimensional spatial tensors. Conversely, it also enables us to 		write quantities invariant under space diffeomorphisms in fully covariant form without the need to introduce new 				propagating degrees of freedom.
	\end{titlepage}

It has been recognized for some time that in order to improve the UV behaviour of the graviton propagator and, thus, the renormalizability of gravity, it is necessary to add higher spatial derivatives to its Lagrangian but no higher time
derivatives. Because this seems to contradict the relativistic local Lorentz
invariance, it was thought necessary to break the symmetry between space and
time. The most notable attempt is the one by Ho\v{r}ava \cite{Horava}, who
constructed a model of quantum gravity with explicitly broken Lorentz symmetry,
which allowed him to add to the action terms dependent on the spatial
Ricci tensor and curvature scalar and their space derivatives (see e.g.
\cite{Wang} and references therein). This is a high price to pay because,
although the Ho\v{r}ava model is renormalizable when projected into the
product space $\mathbb{R}\times\Sigma_{3},$ this property is lost when the model is made covariant by adding one new field \cite{barvinsky}. Various
attempts were made to keep renormalizability of the models while restoring
Lorentz invariance by adding a dynamical scalar or vector \cite{Germani}. Such models exhibit additional propagating degrees of freedom, which limited their
acceptance as a solution to the problem of
renormalizability of gravity.

Mimetic gravity was proposed as a way of separating the scale factor from the
metric and resulted in reproducing Einstein gravity in addition
to half a degree of freedom which could be used to mimic dark matter \cite{mimetic}. The
main observation is that one can define the metric tensor $g_{\mu\nu}$ in
terms of an auxiliary metric $\widetilde{g}_{\mu\nu}$ by the relation%
\begin{equation}
g_{\mu\nu}=\widetilde{g}_{\mu\nu}\left(  \widetilde{g}^{\kappa\lambda}%
\partial_{\kappa}\phi\partial_{\lambda}\phi\right) , \label{1}%
\end{equation}
where $\phi$ is a scalar field. The metric $g_{\mu\nu}$ is invariant under the
scale transformation $\widetilde{g}_{\mu\nu}\rightarrow\Omega^{2}%
\widetilde{g}_{\mu\nu}$ and, as can be easily shown, satisfies the constraint
\begin{equation}
g^{\mu\nu}\partial_{\mu}\phi\partial_{\nu}\phi=1, \label{2}%
\end{equation}
governing the evolution of $\phi$. Thus, instead of introducing the mimetic field $\phi$
through the reparametrization (\ref{1}), it is easier to consider directly the physical metric $g_{\mu\nu}$ together with a constrained scalar field, enforcing (\ref{2}) through a Lagrange multiplier
\cite{Golovnev}. This implies that out of the $11$ variables $g_{\mu\nu}$ and
$\phi$ there are only $10$ independent fields. In the ADM decomposition of $g_{\mu\nu}$,
\begin{equation}
ds^{2}=N^{2}dt^{2}-\gamma_{ij}\left(  dx^{i}+N^{i}dt\right)  \left(
dx^{j}+N^{j}dt\right)  ,\qquad i=1,2,3 \label{3}%
\end{equation}
where $N$ is the lapse function, $N^{i}$ is the shift vector, and $\gamma
_{ij}=-g_{ij}$ is the metric on the spatial 3d hypersurface, the constraint (\ref{2})
can be solved for $N$ in terms of the $10$ variables $N_{i},$ $\gamma_{ij}$
and $\phi$, yielding
\begin{equation}
N^{2}=\frac{\left(  \partial_{0}\phi-N^{i}\partial_{i}\phi\right)  ^{2}%
}{\left(  1+\gamma^{ij}\partial_{i}\phi\partial_{j}\phi\right)  }. \label{4}%
\end{equation}
In the synchronous gauge $N=1,$ $N_{i}=0,$ a solution of
(\ref{2}) is given by
\begin{equation}
\phi=t+A, \label{5}%
\end{equation}
where $A$ is a constant. Since there exists a whole family of synchronous
coordinate systems, corresponding to the freedom of choice of an initial
hypersurface of constant time, this solution is not unique. On the other hand,
$\phi$ can always be used as one particular synchronous time coordinate,
fixing a unique $3+1$ slicing that we will use from now on. The timelike unit
vector $n_{\mu}=\partial_{\mu}\phi$ points in this time direction. In
particular, we can define the projection operator
\begin{equation}
P_{\mu}^{\nu}=\delta_{\mu}^{\nu}-\partial_{\mu}\phi\partial_{\kappa}\phi
g^{\nu\kappa}, \label{6}%
\end{equation}
satisfying the relations
\begin{equation}
P_{\mu}^{\rho}P_{\rho}^{\nu}=P_{\mu}^{\nu},\qquad P_{\mu}^{\nu}\partial_{\nu
}\phi=0. \label{7}%
\end{equation}
In the synchronous slicing from above we have
\begin{equation}
P_{0}^{0}=0,\qquad P_{0}^{i}=0,\qquad P_{i}^{0}=0,\qquad P_{i}^{j}=\delta
_{i}^{j}, \label{8}%
\end{equation}
showing that $P_{\mu}^{\nu}$ projects space-time vectors to space vectors. It
is then clear that in mimetic gravity, using the projection operator and the
vector $n_{\mu}=\partial_{\mu}\phi$, it is possible to construct
four-dimensional tensors whose only non-zero components in the synchronous
gauge are along space directions. For example, as we will show in the following, the expression%
\begin{equation}
\widetilde{R}:=2R^{\mu\nu}\partial_{\mu}\phi\partial_{\nu}\phi-R-\left(
\square\phi\right)  ^{2}+\nabla_{\mu}\nabla_{\nu}\phi\nabla^{\mu}\nabla^{\nu
}\phi\label{9a}%
\end{equation}
coincides with the spatial curvature scalar $^{3}\!R$ of synchronous slices.

In previous works we have shown that in mimetic gravity, without the need to
introduce any additional fields, we can build cosmological models
\cite{mimcos} and solve the singularity problem for Friedmann, Kasner
\cite{Singular} and Black hole \cite{BH} solutions by using the idea of
limiting curvature. More recently we have shown that the idea of asymptotic
freedom can be implemented in mimetic gravity by introducing a $\Box\phi$
dependent effective gravitational constant which vanishes at the limiting
curvature \cite{AF}. Moreover, it was shown that such a dependence does not
introduce higher time derivatives.

The purpose of this letter is to show that within mimetic
gravity we can construct all the terms needed in Ho\v{r}ava gravity using four-dimensional
tensors that reduce to the desired form in the synchronous gauge. We will thus
show that in mimetic gravity it is possible to formulate Ho\v{r}ava gravity in
a diffeomorphism invariant way without introducing ghost-like degrees of freedom.

The basic fields that appear in Ho\v{r}ava gravity are the three-dimensional
tensors and scalars $\kappa_{ij},$ $\kappa,$ ${^{3}\!R}_{ij},$ ${^{3} \!R},$
$D_{k}{^{3} \! R}_{ij},$ and their contractions needed to form space diffeomorphism invariant
expressions. The extrinsic curvature of the synchronous slices $\phi=const.$ is given by
\begin{equation}
\kappa_{ij}=\frac{1}{2}\dot{\gamma}_{ij},\qquad\kappa_{i}^{j}=\gamma
^{jl}\kappa_{il}, \qquad\kappa= \kappa^{i}_{i} = (\ln\sqrt{{\gamma}})^{\cdot},
\end{equation}
where dot denotes $t$ derivative and $\gamma$ is the metric determinant. Using $\phi$, it
can be expressed covariantly as
\begin{equation}
\nabla_{i}\nabla_{j}\phi=-\kappa_{ij},\quad\nabla_{i}\nabla^{j}\phi=\kappa
_{i}^{j},\quad\square\phi=\kappa.
\end{equation}
The non-vanishing components of the four-dimensional Riemann tensor are
determined by
\begin{align}
R_{\: kij}^{0}  &  =D_{i}\kappa_{kj}-D_{j}\kappa_{ki},\\
R_{\: k0j}^{0}  &  =\dot{\kappa}_{jk}-\kappa_{jn}\kappa_{k}^{n},\\
R_{\: kij}^{l}  &  = {^{3}\!R_{\: kij}^{l}}+\kappa_{i}^{l}\kappa_{jk}%
-\kappa_{j}^{l}\kappa_{ik},
\end{align}
where $D_{i}$ and ${^{3}\!R_{\: kij}^{l}}$ are the covariant derivative and the Riemann tensor belonging to the
metric $\gamma_{ij}$. With the help of the above identities, we
can construct the four-dimensional tensor%
\begin{equation}
\widetilde{R}_{\: \rho\mu\nu}^{\sigma}:=P_{\delta}^{\sigma}P_{\rho}^{\gamma
}P_{\mu}^{\alpha}P_{\nu}^{\beta}R_{\: \gamma\alpha\beta}^{\delta}+\nabla_{\mu
}\nabla^{\sigma}\phi\nabla_{\rho}\nabla_{\nu}\phi-\nabla_{\nu}\nabla^{\sigma
}\phi\nabla_{\rho}\nabla_{\mu}\phi
\end{equation}
whose only non-zero components are $^{3}\!R_{\:kij}^{l}$ in the synchronous
gauge. Next, we compute the Ricci tensor components%
\begin{align}
R_{00}  &  =-\overset{.}{\kappa}-\kappa_{ij}\kappa^{ij}\\
R_{0i}  &  =D_{l}\kappa_{i}^{l}-D_{i}\kappa\\
R_{ij}  &  ={^{3}\!R}_{ij}+\kappa\kappa_{ij}-\kappa_{i}^{n}\kappa_{nj}+R_{\:
i0j}^{0}.
\end{align}
These relations allow us to define the tensor%
\begin{equation}
\widetilde{R}_{\mu\nu}:=P_{\mu}^{\alpha}P_{\nu}^{\beta}R_{\alpha\beta}%
+\square\phi\nabla_{\mu}\nabla_{\nu}\phi-\nabla_{\mu}\nabla^{\rho}\phi
\nabla_{\nu}\nabla_{\rho}\phi-R_{\: \mu\delta\nu}^{\gamma}\nabla^{\delta}%
\phi\nabla_{\gamma}\phi,
\end{equation}
whose non-zero components coincide with $^{3}\!R_{ij}$ in the synchronous
gauge. Contracting with $g^{\mu\nu}$, we arrive at (\ref{9a}).

We note in passing that the total derivative $\tfrac{1}{\sqrt{\gamma}}
\partial_{0}\left( \sqrt{\gamma}\kappa\right)$ can be easily eliminated from
the Lagrangian of Einstein-Hilbert gravity, leaving us with
\begin{equation}
-R-2\nabla_{\mu} \left( \square\phi\nabla^{\mu}\phi\right)  = \nabla_{\mu
}\nabla_{\nu}\phi\nabla^{\mu}\nabla^{\nu}\phi- \left( \Box\phi\right)  ^{2} +
\widetilde{R}.
\end{equation}
For manifolds with boundary $\partial M = \left\{  \phi=\phi_{i} \right\}
\cup\left\{  \phi=\phi_{f} \right\} $ consisting of closed spatial
hypersurfaces of constant $\phi$, this has precisely the same effect as adding
the Gibbons-Hawking boundary term.

Space derivatives of the above tensors can be obtained by applying the operator
$P_{\mu}^{\rho}\nabla_{\rho}$. Note that the spatial components of $P_{\rho}^{\gamma}\nabla_{\gamma}\widetilde{R}_{\alpha\beta}$ coincide with $D_{k}{^{3}\!R}_{ij}$ in the synchronous gauge. To obtain a purely spatial tensor, one still must project all four-dimensional indices, i.e. one has to use $P_{\rho}^{\gamma}P_{\mu}^{\alpha}%
P_{\nu}^{\beta}\nabla_{\gamma}\widetilde{R}_{\alpha\beta}.$  Thus, we can now define the analogue of the three-dimensional Cotton tensor
\begin{equation}
{^{3}C}_{j}^{i}=\frac{1}{\sqrt{\gamma}}\epsilon^{ikl}D_{k}\left(  {^{3}%
\!R}_{jl }-\frac{1}{4}\gamma_{jl}\,{^{3}\!R}\right)
\end{equation}
by writing
\begin{equation}
\widetilde{C}_{\nu}^{\mu}:=-\frac{1}{\sqrt{-g}}\epsilon^{\mu\rho\kappa\lambda
}\nabla_{\lambda}\phi\, \nabla_{\rho}\left(  \widetilde{R}_{\nu\kappa
}-\frac{1}{4}g_{\nu\kappa}\widetilde{R}\right) ,
\end{equation}
whose only non-vanishing components in the synchronous gauge are $^{3}
C_{j}^{i}.$

Another object that could be constructed is the Chern-Simons three form
$\omega_{P}$ related to the Pontryagin topological invariant
\begin{align}
R_{\,\rho}^{\sigma}\wedge R_{\,\sigma}^{\rho}  &  =\mathrm{d}\omega_{P}\, ,\\
\omega_{P}  &  =\Gamma_{\mu}^{\nu}\wedge\mathrm{d}\Gamma_{\nu}^{\mu
}+\frac{2}{3}\,\Gamma_{\nu}^{\mu}\wedge\Gamma_{\rho}^{\nu}\wedge\Gamma_{\mu
}^{\rho} \,,
\end{align}
where $\Gamma_{\mu}^{\nu}=\mathrm{d}x^{\rho}\Gamma_{\rho\mu}^{\nu}$ and
$R_{\,\rho}^{\sigma}=\frac{1}{2}R_{\,\rho\mu\nu}^{\sigma}\mathrm{d}x^{\mu
}\wedge\nolinebreak\mathrm{d}x^{\nu}$ are the Christoffel connection one-form
and the curvature two form, respectively. The four-form $\mathrm{d}\phi
\wedge\omega_{P}$ is not parity invariant. Up to a boundary term, its integral
is given by
\begin{equation}
{\int} \mathrm{d}\phi\wedge\omega_{P} = -\int\phi\, R_{\,\rho}^{\sigma}\wedge
R_{\,\sigma}^{\rho}.
\end{equation}
This shows that such a contribution to the action is covariant and invariant
under global shifts of $\phi$. In the synchronous gauge the integrand reduces
to%
\begin{equation}
\epsilon^{ijk}\left(  \Gamma_{i\mu}^{\nu}\partial_{j}\Gamma_{k\nu}^{\mu}%
+\frac{2}{3}\Gamma_{i\nu}^{\mu}\Gamma_{j\rho}^{\nu}\Gamma_{k\mu}^{\rho
}\right)  = {^{3}\omega_{P}}+2\epsilon^{ijk} \kappa_{i}^{n}D_{j}%
\kappa_{kn},
\end{equation}
where
\begin{equation}
^{3}\omega_{P}=\epsilon^{ijk}\left(  \lambda_{im}^{n}\partial_{j}\lambda
_{kn}^{m}+\frac{2}{3}\lambda_{in}^{m}\lambda_{jl}^{n}\lambda_{km}^{l}\right) .
\end{equation}
and $\lambda_{ij}^{ k}$ are the Christoffel symbols calculated for $\gamma_{ij}$. 
The term $2\epsilon^{ijk}\kappa_{i}^{n}D_{j}\kappa_{kn} $ can
be written as
\begin{equation}
\epsilon^{ijk}\nabla_{i}\nabla^{n}\phi R_{njk}^{0}\,,%
\end{equation}
which coincides in the synchronous gauge with
\begin{equation}
\epsilon^{\mu\nu\rho\sigma}\nabla_{\mu}\phi\nabla_{\nu}\nabla^{\lambda}\phi
R_{\lambda\rho\sigma}^{\tau}\nabla_{\tau}\phi.
\end{equation}
Thus, the purely three-dimensional Chern-Simons form can be incorporated in the
action by adding the term%
\begin{equation}%
{\displaystyle\int}
\mathrm{d}\phi\wedge\widetilde{\omega}_{P}:=%
{\displaystyle\int}
\mathrm{d}\phi\wedge\left(  \omega_{P}-\nabla^{\lambda}\mathrm{d}\phi\wedge
R_{\lambda}^{\tau}\nabla_{\tau}\phi\right) .
\end{equation}

All of these manipulations illustrate that any expression invariant under
spatial diffeomorphisms can be written as a combination of four-dimensional tensors that reduces to it in the synchronous gauge.

We conclude by writing an exemplary Ho\v{r}ava action in mimetic gravity, in terms of
four-dimensional tensors and thus completely preserving diffeomorphism
invariance, without the need for new degrees of freedom:%
\begin{align}
S &  =\frac{1}{16\pi G}%
{\displaystyle\int}
\!\mathrm{d}^{4}x\sqrt{-g}\left(  \nabla_{\mu}\nabla_{\nu}\phi\nabla^{\mu
}\nabla^{\nu}\phi-c_{1}\left(  \square\phi\right)  ^{2}+c_{2}\,\widetilde{R}%
\right.  \\
&  \qquad\qquad+c_{3}\,\widetilde{R}^{2}+c_{4}\,\widetilde{R}_{\mu\nu
}\widetilde{R}^{\mu\nu}+c_{5}\,\widetilde{C}_{\nu}^{\mu}\widetilde{C}_{\mu
}^{\nu}+c_{6}\,\eta^{\mu\nu\rho\sigma}\nabla_{\mu}\phi\,(\widetilde{\omega
}_{P})_{\nu\rho\sigma}\nonumber\\
&  \qquad\qquad\left.  +c_{7}\,\eta^{\mu\nu\rho\sigma}\nabla_{\sigma}%
\phi\widetilde{R}_{\mu\alpha}\nabla_{\nu}\widetilde{R}_{\rho}^{\alpha}%
+\textup{\dots}+\lambda\left(  g^{\mu\nu}\partial_{\mu}\phi\partial_{\nu}\phi-1\right)
\right)  ,\nonumber
\end{align}
where $\eta^{\mu\nu\rho\sigma}=\tfrac{1}{\sqrt{-g}}\epsilon^{\mu\nu\rho\sigma
}$. The case
where $c_{1}=c_{2}=1$ and all other couplings vanish is just a rewritten form
of General Relativity with mimetic matter. The constants $c_{1},\dots,c_{7}$ could also be taken as functions of $\square\phi$ in such a way as to reproduce General Relativity in the low curvature limit.

There is no need to repeat calculations done for the Ho\v{r}ava models, as
those could be thought of as a gauge fixed version of a diffeomorphism
invariant theory in the synchronous gauge.

In the projectable Ho\v{r}ava models, the lapse function $N$ is assumed to
depend on time only, $N=N\left(  t\right) .$ These models coincide with the
above family of actions in the synchronous gauge. Their renormalization
analysis was carried out in references \cite{barvinsky}, \cite{barvinsky2},
where they were shown to be renormalizable. When the same analysis was applied
to the non-projectable case where the lapse function is $N=N\left(
x^{i},t\right)$, so that terms dependent on the vector $a_{i}=\frac
{\partial_{i}N}{N}$ can occur, it was found that these models become
non-renormalizable. Attempts were made to construct diffeomorphism invariant
Ho\v{r}ava models by adding a unit vector field $u_{\mu}$, subject to the
hypersurface orthogonality condition $u_{\left[  \mu\right.  }\nabla_{\nu
}u_{\left.  \rho\right]  }=0.$ These models, however, have a spin-1 and spin-0
degree of freedom in addition to the graviton.

The synchronous gauge belongs to the family of temporal gauge, which for
fluctuations of the metric takes the form $n^{\mu}h_{\mu\nu}=0$, where
$g_{\mu\nu}=\eta_{\mu\nu}+h_{\mu\nu}$ and $n_{\mu}=\left(  1,0,0,0\right)  .$
The main advantage of working in this gauge is that the model proposed above
will be power counting renormalizable and that the ghosts associated with
gauge fixing will decouple from the physical S-matrix. The disadvantage is the
need to have an unambiguous prescription for the unphysical singularities of
the form $\left(  q.n\right)  ^{-\alpha}$, $\alpha=1,2$ (cf. \cite{leib}) and the
difficulty in performing higher loop calculations. It is a challenging problem
to perform a detailed analysis of the renormalization program, either in the
synchronous gauge or in a covariant gauge, using the background field method and
integrating out the mimetic constraint, along the lines of \cite{barvinsky}. 
Even though an actual proof could be quite demanding, we expect the mimetic Ho\v{r}ava model presented here to be renormalizable.

\textbf{{\large {Acknowledgments}}}

The work of A. H. C is supported in part by the National Science Foundation
Grant No. Phys-1518371. The work of V.M. and T.B.R. is supported by the
Deutsche Forschungsgemeinschaft (DFG, German Research Foundation) under
Germany's Excellence Strategy -- EXC-2111 -- 390814868.


\begin{thebibliography}{99}                                                                                               %


\bibitem {mimetic}A. H. Chamseddine and V. Mukhanov, \textit{Mimetic Dark
Matter, }JHEP \textbf{1311 }(2013) 135.

\bibitem {Golovnev}A. Golovnev, \textit{On the Recently Proposed Mimetic Dark
Matter, }Phys. Lett. \textbf{B728 }(2014) 39.

\bibitem {mimcos}A. H. Chamseddine, V. Mukhanov and A. Vikman,
\textit{Cosmology with Mimetic Matter, }JCAP \textbf{1406 }(2014) 017.

\bibitem {Singular}A. H. Chamseddine and V. Mukhanov, \textit{Resolving
Cosmological Singularities, }JCAP \textbf{1703 }(2017) 009.

\bibitem {BH}A. H. Chamseddine and V. Mukhanov, \textit{Nonsingular Black
Hole, }Eur. Phys. J. \textbf{C77} (2017) 83.

\bibitem {barvinsky}A. Barvinsky, D. Blas, M. Herrero-Valea, S. Sibiryakov and
C. Steinwachs, \textit{Renormalization of Ho\v{r}ava Gravity, }Phys. Rev.
\textbf{D93 }(2016) 064022.\textit{ }

\bibitem {barvinsky2}A. Barvinsky, M. Herrero-Valea and S. Sibiryakov,
\textit{Towards the renormalization group flow of Ho\v{r}ava gravity in
$(3+1)$ dimensions, }Phys. Rev. \textbf{D100 }(2019) 026012.

\bibitem {Germani}C. Germani, A. Kehagias and K. Sfetsos, \textit{Relativistic
Quantum Gravity at a Lifshitz point, }JHEP \textbf{009 }(2009) 060.

\bibitem {Horava}P. Ho\v{r}ava, \textit{Quantum Gravity at a Lifshitz point,
}Phys. Rev. \textbf{D79 }(2009) 084008.

\bibitem {Wang}A. Wang, \textit{Ho\v{r}ava Gravity at a Lifshitz Point: A
Progress Report, }Int. J. Mod. Phys. \textbf{D26 }(2017) 1730014.

\bibitem {AF}A. H. Chamseddine, V. Mukhanov and T. B. Russ,
\textit{Asymptotically Free Mimetic Gravity, }Eur. Phys. J. \textbf{C79
}(2019) 558.

\bibitem {leib}G. Leibbrandt, \textit{Introduction to noncovariant gauges},
Rev. Mod. Phys. \textbf{59 }(1987) 1067.
\end{thebibliography}
\end{document}